\documentclass[preprint,prd,tightenlines,nofootinbib,floatfix]{revtex4}
\usepackage{graphicx}
\font\symbolfont=cmsy10 at 10 pt
\def\texttilde{{\symbolfont \char'030}}
\def \MSbar {\vbox{\hrule\kern 1pt\hbox{\rm MS}}}
\begin{document}

\title{Next-to-leading order QCD jet production with parton showers
and hadronization}
\author{ Michael Kr\"amer}
\affiliation{Institut f\"ur Theoretische Physik E,
RWTH Aachen, D-52056 Aachen, Germany}
\author{ Stephen Mrenna}
\affiliation{Fermi National Accelerator Laboratory,
Batavia, IL 60510-0500 USA}
\author{ Davison E.\ Soper}
\affiliation{Institute of Theoretical Science, 
University of Oregon, Eugene, OR 97403 USA}
\date{13 September 2005}

\begin{abstract}
  
  We report on a method for matching the next-to-leading order
  calculation of QCD jet production in $e^+e^-$ annihilation with a
  Monte Carlo parton shower event generator (MC) to produce realistic
  final states. The final result is accurate to next-to-leading order
  (NLO) for infrared-safe one-scale quantities, such as the Durham
  3-jet fraction $y_3$, and agrees well with parton shower results for
  multi-scale quantities, such as the jet mass distribution in 3-jet
  events.  For our numerical results, the NLO calculation is matched
  to the event generator {\tt Pythia}, though the method is more
  general.  We compare one scale and multi-scale quantities from pure
  NLO, pure MC, and matched NLO-MC calculations.

\end{abstract}

\maketitle


\section{Introduction}
\label{sec:introduction}

Perturbation theory is the basic tool for deriving predictions for
experiment from the standard model and its extensions. When some of
the particles feel the strong interaction, a perturbative expansion in
powers of the strong coupling $\alpha_s$ can be employed. As long as a
short-distance process is involved, the reasonably small value
($\alpha_s\sim 1/10$) of the running strong coupling justifies such an
expansion, and the leading term is a good first approximation.
However, the estimated error to a leading order (LO) prediction (that
is, the error that one estimates from leaving out all of the higher
order terms) is often so large that one wants a next-to-leading order
(NLO) calculation. In fact, next-to-leading order predictions are
available in the form of computer programs for a wide variety of
processes for electron-positron collisions, electron-hadron
collisions, and hadron-hadron collisions.  The essential limitation is
on the number of partons involved in the hard interaction. Thus $p + p
\to W + 2\ {\it jets} + X$ is available at next-to-leading order while
$p + p \to W + 3\ {\it jets} + X$ is not.

A typical next-to-leading order calculation predicts the expectation
value $\langle {\cal S}\rangle$ of an ``infrared-safe'' observable
${\cal S}$ in the style of an event generator. That is, the program
generates a large number $N$ of events characterized by their final
states $f_n$, with each event having a weight $w_n$. Letting the value
of ${\cal S}$ for state $f_n$ be ${\cal S}(f_{\!n})$, the calculated
expectation value of ${\cal S}$ is
\begin{equation}
\langle {\cal S}\rangle = { 1 \over N}\sum_{n=1}^N
w_n\,{\cal S}(f_{\!n})
\;.
\end{equation}
(A convenient normalization is $(1/N)\sum_{n=1}^N w_n = 1$ 
so that $\langle {1}\rangle = 1$. One could also take 
$\langle {1}\rangle = \sigma_{\rm tot}$.)
The quantity $\langle {\cal S}\rangle$ has the perturbative expansion
\begin{equation}
\langle {\cal S}\rangle = C_0({\cal S})\,\alpha_s^B
+ C_1({\cal S})\,\alpha_s^{B+1} + C_2({\cal S})\,\alpha_s^{B+2} +
\cdots 
\;,
\end{equation}
where B is the power of $\alpha_s$ in the lowest-order graph for
$\langle {\cal S}\rangle$. The numerical result of the NLO calculation
is not exact, but produces the first two terms of this expansion.

Typical programs that do next-to-leading order calculations suffer
from a serious deficiency: the final states $f_n$ are not realistic.
First of all, they consist of just a few partons. In the case of $e^+
+ e^- \to 3\ {\it jets}$, which we consider in this paper, the final
states consist of a quark, an antiquark, and one gluon or else a
quark, an antiquark, and two gluons or another quark-antiquark pair.
This is not the kind of state that can be used directly as input to a
detector simulation.

\begin{figure}[tb]
\vspace*{-10mm}
\includegraphics[width = 13 cm]{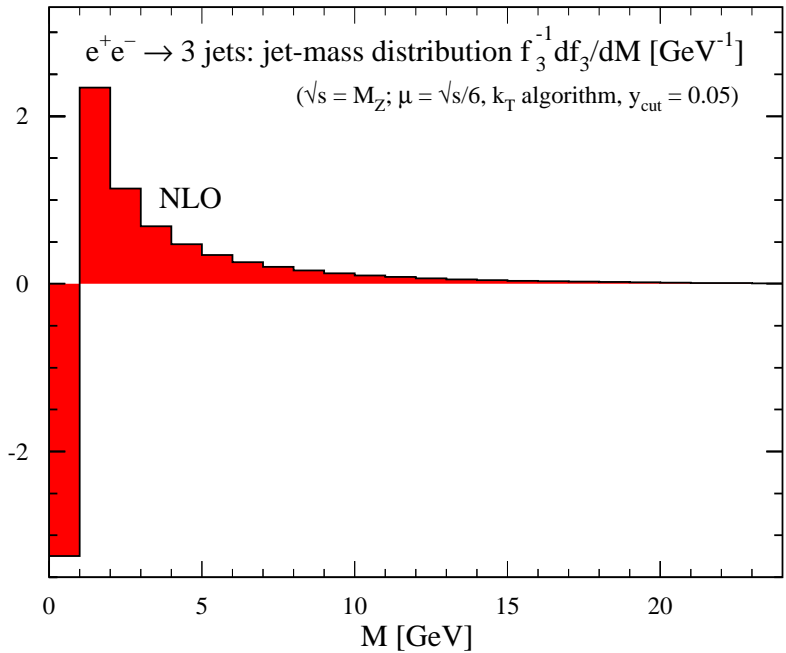}
\vspace*{-7mm}
\caption{Jet mass distribution in three-jet events, $f_3^{-1}\,df_3/dM$,
  calculated at next-to-leading order for $\sqrt s = M_Z$. The three
  jets are identified using the $k_T$ algorithm with $y_{\rm cut} =
  0.05$. Then $M$ is the invariant mass of one of the three jets, with
  each event making three contributions to $df_3/dM$. The
  renormalization scale is chosen as $\mu = \sqrt s / 6$ and
  $\alpha_s(M_Z)=0.118$. This is a pure NLO calculation
  using~\cite{beowulf}.  There is a large negative contribution in the
  first bin.  }
\label{fig:df3dmNLO}
\end{figure}

\begin{figure}[tb]
\vspace*{-10mm}
\includegraphics[width = 13 cm]{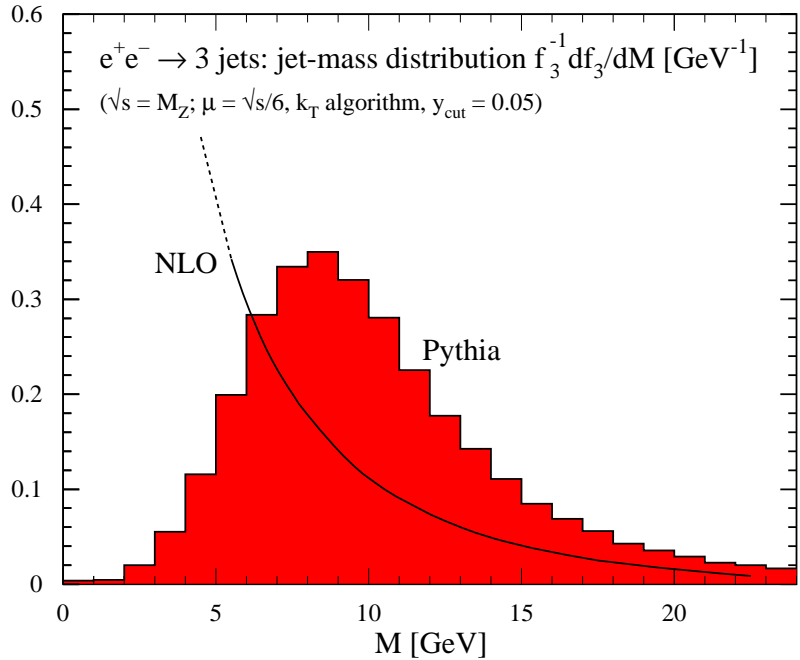}
\vspace*{-7mm}
\caption{Jet mass distribution in three-jet events, $f_3^{-1}\,df_3/dM$,
  calculated according to {\tt Pythia} \cite{Pythia} with default
  parameters for $\sqrt s = M_Z$. The cross section is defined as in
  Fig.~\ref{fig:df3dmNLO}. The jet mass distribution from the
  perturbative NLO calculation in Fig.~\ref{fig:df3dmNLO} is shown for
  comparison.  }
\label{fig:df3dmPythia}
\end{figure}

The situation is even worse than might be deduced from the fact that
the final states consist of few partons instead of (many) hadrons.  At
its core, an NLO calculation depends on the use of an infrared-safe {\it
  inclusive} observable since it depends on a cancellation of
divergences associated with different numbers of on-shell partons. As
a concrete example, consider the fraction of events in $e^+ + e^- \to
{\it hadrons}$ that contain precisely three jets as determined by the
$k_T$ (or Durham) algorithm~\cite{kTjets} with $y_{\rm cut} = 0.05$.
Call this fraction $f_3$. We have
\begin{equation}
\langle f_3\rangle = C_0(f_3)\,\alpha_s
+ C_1(f_3)\,\alpha_s^{2} + C_2(f_3)\,\alpha_s^{3} + \cdots
\;.
\end{equation}
A next-to-leading order program can accurately predict $C_0$ and
$C_1$. Now suppose that we use the program to determine the
differential three-jet fraction $df_3/dM$, where $M$ is the mass of
one of the jets (so that each event contributes three entries for
$df_3/dM$ and $\int \!dM\,df_3/dM = 3 f_3$). 
Even if one is ultimately interested in $f_3$, it is
reasonable to be interested in predicting this more detailed quantity,
because a realistic detector may respond differently to narrow and
wide jets and we may be concerned with how much detector effects
influence the measurement.\footnote{ Actually, for this purpose one
  would be more interested in $df_3/dM_1\,dM_2\,dM_3$, but we discuss
  $df_3/dM$ for reasons of simplicity.}  Now, one can expect the NLO
result to be approximately correct for $df_3/dM$ as long as $M$ is large, but
for $M/\sqrt s \ll 1$, a sophisticated user will expect a pure NLO
program to give an answer that is not physically realistic. We
illustrate what goes wrong by simply plotting the result for
$f_3^{-1}\,df_3/dM$ versus $M$ from a purely NLO program in
Fig.~\ref{fig:df3dmNLO}.  We see that $f_3^{-1}\,df_3/dM$ becomes
large for small $M$ and that the fraction of events in the first bin
is large and negative. As $M \to 0$, $df_3/dM$ according to the NLO
calculation develops a $\log(M)/M$ singularity together with a term
proportional to $\delta(M)$ with an infinite negative coefficient. The
integral $f_3 = (1/3) \int\!dM\, df_3/dM$ of this highly singular function
gives an accurate estimate of the three-jet fraction $f_3$ but the
differential distribution is completely unphysical.  In fact, it
requires a binning of the distribution even to get finite numbers.  A
much more realistic expectation for $f_3^{-1}\,df_3/dM$ can be
obtained by using a parton shower Monte Carlo event
generator~\cite{Pythia,Herwig,Ariadne}.  Recall that the shower Monte
Carlo programs add up a series of logarithmic terms, approximating the
behavior of Feynman graphs near the singularities corresponding to
collinear parton splitting or soft gluon emission.  The prediction of
the shower Monte Carlo {\tt Pythia}~\cite{Pythia} for
$f_3^{-1}\,df_3/dM$ is shown in Fig.~\ref{fig:df3dmPythia} and is
compared to the NLO calculation. We see that the NLO perturbative
calculation agrees qualitatively with {\tt Pythia} for large $M$, 
but fails for small $M$.

Clearly if one were interested in how the differential response of a
detector to wide and narrow jets would affect a measurement of $f_3$,
one would be better off using {\tt Pythia} than the NLO program, even
though {\tt Pythia} has only leading-order accuracy for $f_3
= (1/3) \int\!dM\, df_3/dM$.  However, would it not be better to add parton
showering and hadronization according to {\tt Pythia} to the NLO
calculation? That is the topic discussed in this paper.

There has been substantial recent progress in adding a parton shower
and hadronization to an NLO calculation. Frixione, Nason and Webber
\cite{FrixioneWebberI, FrixioneWebberII, FrixioneWebberIII,
  Nason:2004rx} have presented and implemented an algorithm that makes
use of the shower Monte Carlo program {\tt Herwig}~\cite{Herwig} to do
NLO calculations for a selection of processes for which there are two
hadrons in the initial state and at leading order the final state
particles that carry color are massive. Processes available include
the hadroproduction of single vector and Higgs bosons, vector boson
pairs, heavy quark pairs, and lepton pairs. Two of the present authors
have described an algorithm for a process with massless color carrying
particles in the final state, namely $e^+ + e^- \to 3\ {\it jets}$
\cite{nloshowersI,nloshowersII}.  This algorithm was implemented with
an NLO calculation coupled to a small self-contained shower generator
with no hadronization.

The calculations of Refs.~\cite{nloshowersI,nloshowersII} sufficed to
demonstrate the principles of the algorithm, but for practical
purposes one would like to have a shower algorithm that has been
tested against data (as in {\tt Herwig} and {\tt Pythia}) and one
would like to include hadronization. For this reason, we present here
results from coupling the program described in
Refs.~\cite{nloshowersI,nloshowersII} to the standard shower Monte
Carlo program {\tt Pythia}, including its hadronization following the
Lund string model. The code described here is available at the site
\cite{beowulfcode}. Other work on this general subject can be found in
Refs.~\cite{otherwork}.

\section{Sketch of the algorithm}
\label{sec:algorithm}

The general algorithm that we use to couple an NLO program for $e^+ +
e^- \to 3\ {\it jets}$ to a shower Monte Carlo is described in some
detail in Refs.~\cite{nloshowersI,nloshowersII}, so we present just a
sketch here. Specific details about the coupling to {\tt Pythia} that
go beyond Refs.~\cite{nloshowersI,nloshowersII} are described in the
following section.

\begin{figure}
\vspace*{5mm}
\includegraphics[width = 6 cm]{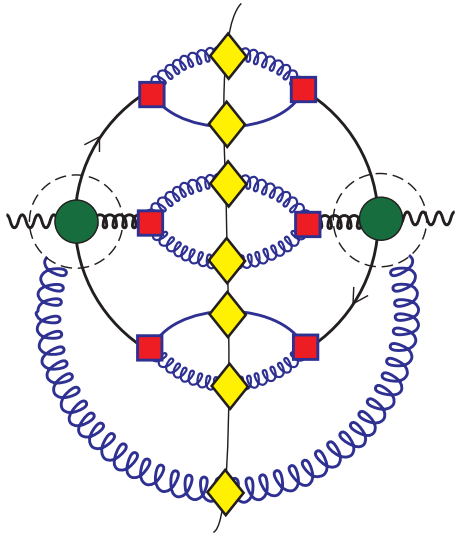}
\caption{Parton splitting in
  Refs.~\protect\cite{nloshowersI,nloshowersII}.  The filled circles
  represent graphs for the Born amplitude and complex conjugate
  amplitude.  Each of the partons emerging from the Born amplitude
  splits into two partons with a vertex, represented by the squares,
  that includes a Sudakov suppression factor. The extra gluon coming
  from the circles on the right and left represents the soft gluon
  radiated from the three jets. Each of the seven daughter partons
  undergoes further, secondary, splittings according to {\tt Pythia}
  and enters the final state as a complete shower with hadronization.
  The secondary splittings are represented by the diamonds.}
\label{bornshower}
\end{figure}

Consider first the Born graphs. A quark-antiquark-gluon state is
generated with a weight proportional to one of the Born graphs times
the complex conjugate of one of the Born graphs. Each of these partons
splits into two with a probability described by a splitting function
and a Sudakov exponential that represents the probability not to have
split at a higher virtuality. This splitting is called a primary
splitting and is represented by the square vertices in
Fig.~\ref{bornshower}. The splitting functions have the right
singularities to represent the collinear limit of QCD matrix elements
and the collinear $\times$ soft limit, but they are not correct in the
limit of the emission of a wide angle soft gluon. Accordingly, we
generate a gluon emission with a weight that matches the matrix
element for the radiation of a very soft gluon from the antenna
produced by the three outgoing partons. This emission is represented
by the gluon in Fig.~\ref{bornshower}, which is drawn to suggest that
it is emitted from the outgoing partons as a whole but not from any
particular one of them.

Next, the seven parton final state thus generated is modified somewhat
as described in the following section and fed to {\tt Pythia}, which
creates a complete shower and hadronization. We have modified {\tt
  Pythia} slightly so that it is able to accept a partially developed
shower and generate suitably narrow jets from each of the seven
partons.  We outline the required modifications in the following
section.

\begin{figure}
\vspace*{7mm}
\includegraphics[width = 9 cm]{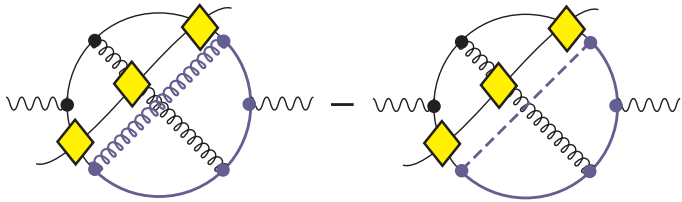}
\caption{Treatment of order $\alpha_s^{B+1}$ graphs. One particular cut
  diagram is shown, in this case a cut diagram with a virtual loop.
  The second diagram illustrates a subtraction term, which corresponds
  to a contribution at order $\alpha_s^{B+1}$ from
  Fig.~\ref{bornshower}. The partonic final state is generated with a
  weight proportional to the difference of matrix elements, then this
  final state is sent to {\tt Pythia}, which is represented by the
  diamonds.}
\label{NLOshower}
\end{figure}

The splitting incorporated in the Born graphs has two important
features. First, when the virtuality $\bar q^2$ of the pair of
daughter partons is small, the splitting probability is proportional
to $(P(x)/\bar q^2) d\bar q^2 dx$ where $x$ is the share of the
momentum carried by one of the daughter partons and $P(x)$ is the
appropriate Altarelli-Parisi splitting function. Second, the collinear
singularity at $\bar q^2 \to 0$ is damped by a Sudakov factor with the
behavior $\exp(-\alpha_s\, c\log^2(\bar q^2))$ for $\bar q^2 \to 0$.
Thus each of the three original partons makes a jet, but in the limit
of small $\alpha_s$, each jet is usually very narrow and appears in an
infrared-safe measurement like a single massless parton.  There is a
qualifying adverb ``usually'' here. A fraction $\alpha_s$ of the time,
one of the splittings has a substantial virtuality and we get a
four-jet final state. Thus there is an order $\alpha_s^{B+1}$ effect
in which some probability is removed from the three-jet final state
and given to a four-jet final state. In a purely NLO calculation, this
effect is included in the $\alpha_s^{B+1}$ graphs. Since we do not
want any part of the order $\alpha_s^{B+1}$ contribution to be counted
twice, we need to subtract these terms from the $\alpha_s^{B+1}$
graphs. One of these subtraction terms is illustrated in
Fig.~\ref{NLOshower}. The subtraction terms have the effect of
removing collinear and soft divergences from the order
$\alpha_s^{B+1}$ graphs.

In addition to the Born graphs, there are subtracted order
$\alpha_s^{B+1}$ graph with either three or four partons in the final
state. Each of the final state partons from these graphs is fed to
{\tt Pythia}, which creates a complete shower and hadronization using
appropriate initial conditions as described in the following section.
This showering is represented by the diamonds in Fig.~\ref{NLOshower}.
 
The Sudakov suppression factors play an important role in the primary
showering that takes place before the partons are passed to {\tt
  Pythia}. In the case of quark splitting, the Sudakov factor has the
form
\begin{displaymath}
\exp\left(-\int_{\bar q^2}^\infty \frac{d\bar l^2}{l^2} \int_0^1 dz\
\frac{\alpha_s}{2\pi}\,{\cal P}_{g/q}(\bar l^2,z,|\vec q^{\,2}|)
\right)\;,
\end{displaymath}
where $\bar q^2$ is the virtuality of the splitting and $\vec q$ is the
momentum of the parent quark. The function
${\cal P}_{g/q}(\bar l^2,z)$ represents ``virtual'' splittings and is
derived from the one loop quark self-energy graph in the Coulomb
gauge.  It reduces to the usual Altarelli-Parisi splitting function in
the limit $\bar l^2 \to 0$ at fixed $z$. However the usual $1/z$
singularity is absent, being cut off for $z < \bar l^2/|\vec
q^{\,2}|$, where $\vec q$ is the momentum of the mother quark. The
integration over $z$ produces a logarithm of $\bar l^2/|\vec
q^{\,2}|$.  Then the integration over $\bar l^2$ produces a second
logarithm so that splitting at very small $\bar q^2$ is suppressed by
a Sudakov factor that is, in a first approximation, $\exp(-\alpha_s\,
c\log^2(\bar q^2))$.  In Refs.~\cite{nloshowersI,nloshowersII}, the
Sudakov factor is not derived as a summation of logarithms. Rather, it
is derived as a useful way of expressing the effect of virtual graphs
on small $\bar q^2$ splittings. Compare this to the standard
perturbative treatment of small $\bar q^2$ splittings in which large
positive probabilities from real parton emissions are counterbalanced
by large negative probabilities from virtual parton graphs. With the
Sudakov factor, we avoid the small $\bar q^2$ singularities: we
suppress them with an exponential factor. When expanded in powers of
$\alpha_s$, the result is the same as given by the standard
perturbative calculation.  Next, one simply notes that the Sudakov
suppression factor is just what one wants as the first step in a
shower Monte Carlo algorithm, where this factor appears as a way of
summing logarithms.

This algorithm is based on the order $\alpha_s^B$ and $\alpha_s^{B+1}$
graphs for $e^+e^- \to 3\ {\it jets}$. It is designed to produce NLO
calculations for infrared-safe three-jet observables while, at the
same time, providing a reasonably accurate parton shower description
for the inner structure of the three jets. Future NLO-MC hybrid
programs will likely have a mechanism for switching from a three-jet
description to a two-jet description, as in Refs.~\cite{Catani:2001cc}
and \cite{Mrenna:2003if} at lowest order and Ref.~\cite{Nagy:2005aa}
at NLO.  Such a mechanism is not built into the present program. For
this reason, the present program should not be used to calculate
observables that receive significant contributions from two-jet final
states. In fact, in the program, events that are too close to a
two-jet configuration based on the thrust of the perturbative event
are simply cut from the calculation.

\section{Merging with {\tt Pythia}}
\label{sec:pythia}

In this section, we address the issue of how to pass the perturbative
events after primary showering through a standard event generator to
include additional jet structure and hadronization.  In this paper, we
use the {\tt Pythia} event generator, though there is no limitation
(in principle) to using another one.

For some time now, {\tt Pythia} has been able to perform parton
showering and hadronization on externally-generated partonic
configurations.  This has become technically simpler with the adoption
of a general data structure by several Monte Carlo authors
\cite{lha1}.  The parton shower that is added, however, is based on
partons that are generated from a hard process at a single scale. We
need something different since the further showering of the partonic
state sent to {\tt Pythia} needs to recognize several scales that were
involved in generating these partons.

We begin with seven partons from a Born graph as depicted in
Fig.~\ref{bornshower}, three partons from an order $\alpha_s^{B+1}$
graph with a virtual loop as depicted in Fig.~\ref{NLOshower}, or four
partons from an order $\alpha_s^{B+1}$ graph with no virtual loop.
These final state partons are generated with flavor identifications
$\{g,u,\bar u, d, \dots\}$ and with labels indicating color
connections based on the color structure of the underlying Feynman
graphs. (The idea of color connections as used in parton shower Monte
Carlo programs neglects terms suppressed by $1/N_{\rm c}^2$, where
$N_{\rm c}$ is the number of colors, and neglects quantum interference
graphs. Thus the assignment of color connections can only be a rough
approximation in a calculation with exact color factors for the graphs
and with quantum interference included. Nevertheless, an assignment of
color connections is needed in order for {\tt Pythia} to generate
string hadronization.) We also supply the quark masses used in {\tt
  Pythia} to the final state quarks. The momenta for final state
partons were generated in the approximation that all partons are
massless, but the momenta are adjusted so that final state quarks are
on-shell with the proper masses.

Before the partonic final state is passed to {\tt Pythia}, a certain
amount of analysis is needed.

The first step is to account for the minimum virtuality of 1 ${\rm
  GeV}^2$ allowed in a {\tt Pythia} shower. In contrast, there are no
infrared cutoffs in the virtualities of the parton splittings in
Fig.~\ref{bornshower} or in the energy of the soft gluon there. Thus
in order to generate partonic states that are sensible within the
context of {\tt Pythia}, the program checks whether any of the three
splittings indicated in Fig.~\ref{bornshower} had a virtuality less
than the minimum. If there was an unallowed splitting, the two
daughter partons are replaced by the mother parton and the parton is
flagged so that it is not showered by {\tt Pythia}. In addition, if
the soft gluon in Fig.~\ref{bornshower} (or any other final state
gluon that did not arise from a $1 \to 2$ splitting) has an energy
less than 100 MeV, the gluon is erased and its momentum and color is
reassigned to the other partons.  This step leaves us with seven or
fewer partons to be passed to {\tt Pythia}.

The next step is to generate a synthetic shower history that could
have resulted in the final state partons. This shower history is
generated based on the $k_T$-algorithm modified to account for the
flavors and colors of the final state partons, so that we get a shower
history that is allowed in QCD at leading order in $1/N_{\rm c}^2$. In
generating this shower history, we define the resolution variable to
be the {\tt LUCLUS} resolution variable that is used in {\tt Pythia}:
\begin{equation}
k_{T,ij} =  2 (1 - \cos \theta_{ij} )\
\frac{E_i^2 E_j^2}{(E_i + E_j)^2}
\;.
\end{equation}
Here the energies and angles are defined in the overall 
center-of-mass frame of the event.
The jet algorithm examines all of the $k_{T,ij}$ values for pairs
$\{i,j\}$ of partons that have flavors and color connections that
would allow them to be combined. The pair with the smallest $k_{T,ij}$
{\it is} combined (by adding their four-momenta). Then the algorithm
repeats until, at last, we have just one remaining quark-antiquark
pair. We use this synthetic shower history to define, for each final
state parton $i$, a certain initial virtuality scale $Q_i$, a maximum
transverse momentum $k_{T,i}^{\rm max}$, and a maximum angle
$\theta_i^{\rm max}$. The first splitting generated by {\tt Pythia}
for parton $i$ should have $k_T < k_{T,i}^{\rm max}$ and $\theta <
\theta_i^{\rm max}$.

To define $k_{T,i}^{\rm max}$, we find the $k_T$ of the splitting that
produced parton $i$ in the synthetic shower history. We define
$k_{T,i}^{\rm max}$ to be the lesser of this and a global maximum
$k_T$, which is the largest $k_{T,ij}$ in the shower. The purpose of
using this global scale is to guarantee that the hardest interaction
is generated in the perturbative NLO calculation, not by {\tt Pythia}.

To define $\theta_i^{\rm max}$, we find the one or two final state
partons $j$ with which parton $i$ is color connected and define
$\theta_i^{\rm max} = \min_j \theta_{ij}$.

To define the initial virtuality scale $Q_i$ there are a number of
cases.  When parton $i$ is a quark or antiquark that can be traced
back to the $\gamma^{*}/Z$ vertex, we define $Q_i^2 = s$. When parton
$i$ is a quark or antiquark that can be traced back to a $g \to q \bar
q$ vertex, we define $Q_i^2 = p_g^2$ where $p_g$ is the four-momentum
of the gluon at which the quark or antiquark line started. When parton
$i$ is a gluon that originated at a $q \to q g$ or $\bar q \to \bar q
g$ vertex, we define $Q_i^2 = p_{\rm m}^2$ where $p_{\rm m}$ is the
four-momentum of the quark or antiquark that is the mother for the
gluon. When parton $i$ is a gluon that originated at a $g \to g g$
vertex and is the less energetic of the two sister gluons, we define
$Q_i^2 = p_{\rm m}^2$ where $p_{\rm m}$ is the four-momentum of the
mother gluon.  When parton $i$ is a gluon that originated at a $g \to
g g$ vertex and is the more energetic of the two sister gluons, we
define $Q_i^2 = p_{\rm mm}$ where $p_{\rm mm}$ is the four-momentum of
the grandmother of the gluon.

The idea now is to ask {\tt Pythia} to generate further showering for
each final state parton $i$ according to the standard {\tt Pythia}
splitting kernels, but with the first splitting limited by the
requirements $k_T < k_{T,i}^{\rm max}$ and $\theta < \theta_i^{\rm
  max}$.  In order to incorporate the $k_T$ restriction into a shower
based on the initial virtuality scale $Q_i$, we follow the procedure
of \cite{Mrenna:2003if}, based on the suggestion of
\cite{Catani:2001cc}.  The shower for each parton $i$ is started at
scale $Q_i$ but splittings that do not satisfy $k_T < k_{T,i}^{\rm
  max}$ are ignored (``vetoed''), where $k_T$ is the parton shower
definition of this quantity, ${k_{T}^2} = z(1-z)m^2$, using
$z$ as the energy fraction of one daughter with respect to the mother
of virtuality $m$.  Splittings that do not satisfy $\theta <
\theta_i^{\rm max}$ are also vetoed, as is standard in {\tt Pythia}.
Once a first allowed splitting has been generated, the rest of the
shower from parton $i$ is generated according to the normal algorithms
of {\tt Pythia}. After parton showering, {\tt Pythia} generates
hadrons by string fragmentation.

\section{Results}
\label{sec:results}

In this section, we test the program that combines the NLO calculation
with showers and hadronization as sketched in
Secs.~\ref{sec:introduction} and \ref{sec:algorithm}. We shall refer
to this program as {\tt NLO+PS+Had}. If we use just the Born graphs
with showers and hadronization, leaving out the contributions from
order $\alpha_s^2$ graphs, we shall refer to the program as {\tt
  LO+PS+Had}.  For comparison, we also display results from a pure
lowest order perturbative calculation, {\tt LO}, a pure
next-to-leading order perturbative calculation, {\tt NLO}, and just
the {\tt Pythia} program (version 6.221 with default parameters unless
specially noted).  Our standard choice for the c.m.\ energy is $\sqrt
s = M_Z$ with $\alpha_s(M_Z)=0.118$. In the perturbative calculations,
we need to choose a renormalization scale.  Our default choice is $\mu
= \sqrt s / 6$, based on the observation that in jet production for
hadron physics the choice $\mu = E_T/2$ works well and in the $e^+e^-$
case each jet has a typical energy $E \approx \sqrt s/3$.

We will first examine the three-jet fraction $f_3$ as a function of
the jet resolution parameter $y_{\rm cut}$. Here the hope is that
$f_3[\mbox{\tt NLO+PS+Had}]$ will match $f_3[{\tt NLO}]$ reasonably
well since $f_3[\mbox{\tt NLO+PS+Had}]$ is supposed to be correct to
next-to-leading order. Then we will examine $f_3^{-1}\,df_3/dM$ at a
fixed choice for $y_{\rm cut}$, which we take as $y_{\rm cut} = 0.05$
(chosen because at this value $f_3$ is approximately
$1\times\alpha_s$). For $f_3^{-1}\,df_3/dM$ we recall that the pure
NLO result is completely unrealistic. The hope is that
$f_3^{-1}\,df_3[\mbox{\tt NLO+PS+Had}]/dM$ will match
$f_3^{-1}\,df_3[{\tt Pythia}]/dM$ reasonably well since {\tt Pythia}
is generally known to give a pretty realistic match to data.

\begin{figure}[tb]
\vspace*{-10mm}
\includegraphics[width = 13 cm]{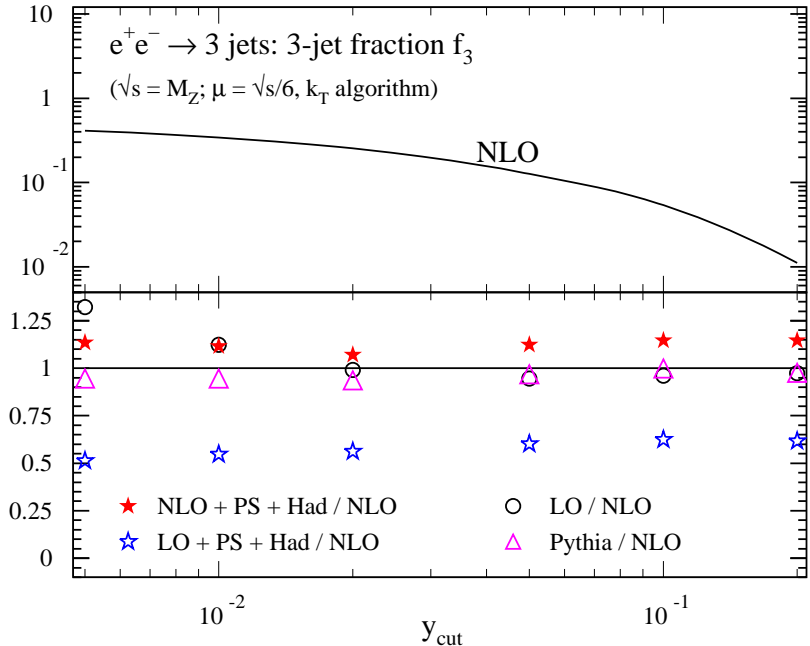}
\vspace*{-7mm}
\caption{
  Three-jet fraction, $f_3$, versus $y_{\rm cut}$. The top panel shows
  $f_3[{\tt NLO}]$. The bottom panel shows the ratios of $f_3[{\tt
    NLO+PS+Had}]$, $f_3[{\tt LO+PS+Had}]$, $f_3[{\tt LO}]$, and
  $f_3[{\tt Pythia}]$ to $f_3[{\tt NLO}]$. The parameters are as in
  Figs.~\ref{fig:df3dmNLO} and \ref{fig:df3dmPythia}.}
\label{fig:f3vsycut}
\end{figure}

We begin with $f_3$ as a function of the jet resolution parameter
$y_{\rm cut}$, which we display in Fig.~\ref{fig:f3vsycut}. We first
note that with our choice of renormalization scale, $f_3[{\tt LO}]$ is
rather close to $f_3[{\tt NLO}]$. This agreement can be taken as a
sign that the choice or renormalization scale was sensible.

Next, we note that $f_3[\mbox{\tt LO+PS+Had}]$ is quite a lot smaller
than $f_3[{\tt NLO}]$ or $f_3[{\tt LO}]$. What has happened is that
the first stage of showering in the algorithm used here changes the
normalization by an amount proportional to $\alpha_s$. This change is
not as alarming as it might seem. For $y_{\rm cut} = 0.05$ we have
$f_3[\mbox{\tt LO+PS+Had}]/f_3[{\tt LO}] \approx 0.5$. If we write
this as one factor for each jet,
\begin{equation}
\frac{f_3[\mbox{\tt LO+PS+Had}]}{f_3[{\tt LO}]} = 
\Big[1 - X\,\alpha_s(\sqrt s / 6)\Big]^3
\end{equation}
with $\alpha_s(\sqrt s / 6) \approx 0.16$, then the coefficient $X$ is
not large, $X \approx 1.5$. One could make $X$ closer to zero by
modifying the functions used in the primary showering, as discussed in
the last paragraph of Sec.~IV of Ref.~\cite{nloshowersI}. For instance
one could make the Sudakov exponent smaller. However, we do not
attempt such an adjustment in this paper.

We are now ready to look at $f_3[\mbox{\tt NLO+PS+Had}]$, for which
the correction terms based on order $\alpha_s^2$ graphs are designed
to produce a result that is correct to order $\alpha_s^2$. Indeed, we
see that $f_3[\mbox{\tt NLO+PS+Had}]$ matches $f_3[{\tt NLO}]$ to
about 10\%, which is within the error that one expects for a
next-to-leading order calculation.

Finally in Fig.~\ref{fig:f3vsycut} we show $f_3[{\tt Pythia}]$. We see
that $f_3[{\tt Pythia}]$ matches $f_3[{\tt NLO}]$ quite well, even
though {\tt Pythia} does not contain all the terms needed for
next-to-leading order accuracy. On the other hand, {\tt Pythia} does
contain {\it some} next-to-leading order terms in the form of its
choice of scale and in the generation of showers. Evidently, these
terms do quite a good job of approximating the full next-to-leading
order result. In fact, one might be inclined to rely entirely on {\tt
  Pythia} if it were not for the fact that one cannot really tell how
accurate its approximations are without having a full next-to-leading
order calculation with which to compare it.

\begin{figure}[tb]
\vspace*{-10mm}
\includegraphics[width = 13 cm]{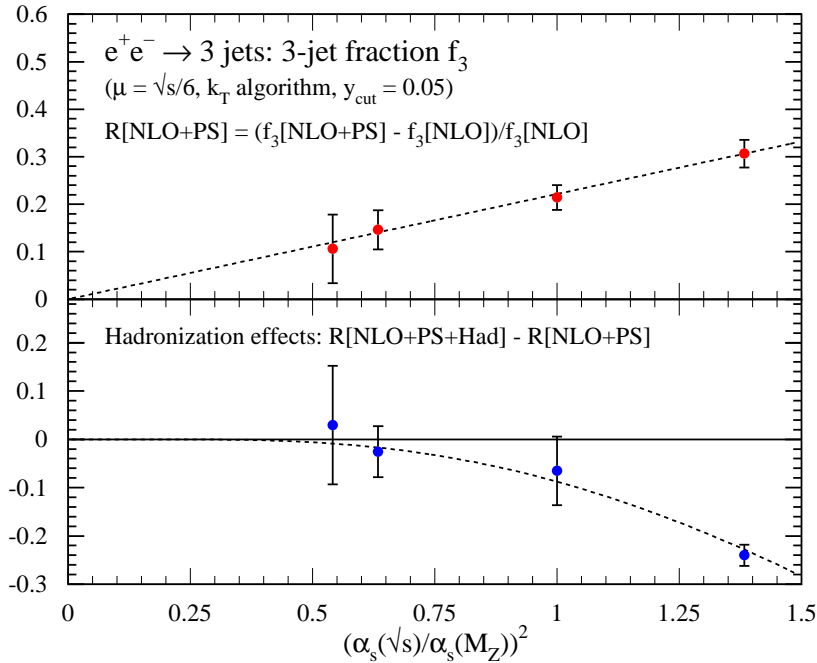}
\vspace*{-7mm}
\caption{
  Comparison of the NLO calculation with showers to a pure NLO
  calculation.  In the top panel, we plot the ratio $R[\mbox{\tt
    NLO+PS}]$ defined in Eq.~(\ref{Rdef}). We take $y_{\rm cut}= 0.05$
  and $\mu = \sqrt s/6$. The ratio $R$ is calculated for the c.m.\ 
  energies $\sqrt s = 34$~GeV, $M_Z$, 500~GeV and 1000~GeV,
  corresponding to $\alpha_s = 0.139$, 0.118, 0.0940 and 0.0868,
  respectively, and is plotted versus $\alpha_s^2(\sqrt{s})
  /\alpha_s^2(M_Z)$. For comparison, the curve $R = 0.22\,\alpha_s^2
  (\sqrt{s})/\alpha_s^2(M_Z)$ is shown. In the bottom panel, we show
  the difference between $R$ calculated with hadronization and $R$
  calculated without hadronization. For comparison, we show the curve
  $\Delta R = 8\ {\rm GeV}/\sqrt s$.}
\label{fig:R}
\end{figure}

A more rigorous perturbative test of whether {\tt NLO+PS}, without
hadronization in this case, is indeed correct to next-to-leading order
can be obtained by studying the ratio
\begin{equation}
R[\mbox{\tt NLO+PS}] = \frac{f_3[\mbox{\tt NLO+PS}] 
 - f_3[\mbox{\tt NLO}]}{f_3[\mbox{\tt NLO}]}
\;.
\label{Rdef}
\end{equation}
The difference between $f_3[\mbox{\tt NLO+PS}]$ and $f_3[\mbox{\tt
  NLO}]$ should be of order $\alpha_s^{3}$ so that the ratio $R$
should have a perturbative expansion that begins at order
$\alpha_s^2$. We can test this by evaluating $f_3$ at different c.m.\ 
energies $\sqrt s$ and plotting the ratio $R$ against
$\alpha_s^2(\sqrt{s})$. The results of this test are shown in the
upper panel of Fig.~\ref{fig:R} for the c.m.\ energies $\sqrt s =
34$~GeV, $M_Z$, 500~GeV and 1000~GeV, corresponding to $\alpha_s =
0.139$, 0.118, 0.0940 and 0.0868, respectively. We see the expected
shape of the $R$ curve, which approaches a straight line through the
origin as $\alpha_s^2 \to 0$. For comparison, we show the line $R =
0.22\, \alpha_s^2(\sqrt{s})/\alpha_s^2(M_Z)$. In the lower panel of
Fig.~\ref{fig:R} we display the impact of hadronization by plotting
the difference $\Delta R = R[\mbox{\tt NLO+PS+Had}] - R[\mbox{\tt
  NLO+PS}]$. We observe that hadronization corrections reduce $f_3$ by
about 20\% at $\sqrt{s} = 34$~GeV and by about 5\% at $\sqrt{s} =
M_Z$.  They are negligible for larger c.m.\ energies. For comparison,
we show the curve $\Delta R = 8\ {\rm GeV}/\sqrt s$. The fact that the
curve is similar to the data suggests that the combined NLO-MC program
gives power law hadronization effects.

\begin{figure}[htb]
\vspace*{-10mm}
\includegraphics[width = 13 cm]{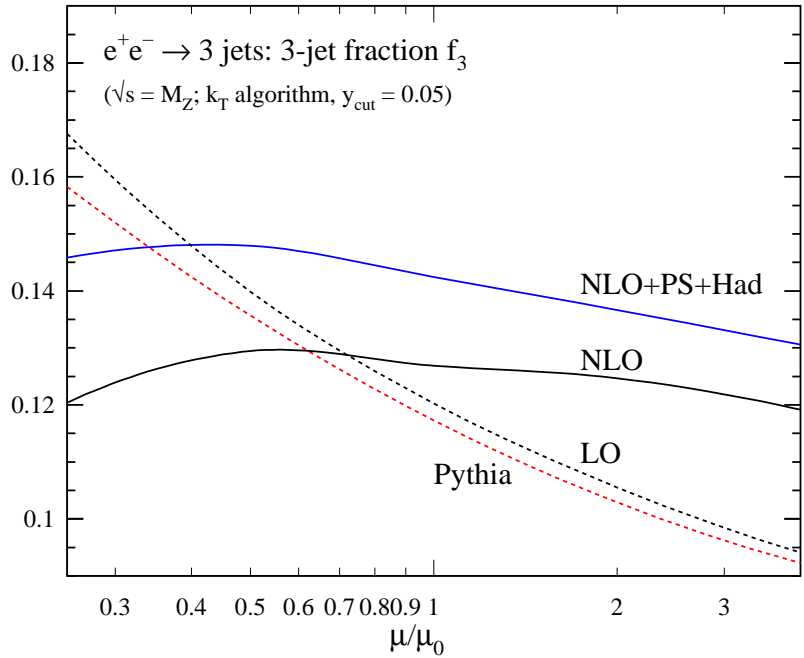}
\vspace*{-7mm}
\caption{
  Three-jet fraction, $f_3$ versus $\mu/\mu_0$, where $\mu$ is the
  renormalization scale and $\mu_0$ is its default value.  We show the
  result of the full {\tt NLO+PS+Had} calculation and for the pure
  {\tt NLO} calculation. In these calculations the default scale is
  $\mu_0 = \sqrt s/6$. For comparison, we also show the result of
  varying the scale in plain {\tt Pythia}. In this case, the default
  value, $\mu_0$ of the scale $\mu$ in $\alpha_s(\mu)$ is the
  transverse momentum in the parton splitting. The ratio $\mu/\mu_0$
  is the inverse of the {\tt Pythia} parameter {\tt PARJ(81)}. We
  choose default parameters as in Fig.~\ref{fig:df3dmNLO}.}
\label{fig:f3vsmu}
\end{figure}

A next-to-leading order calculation is supposed to do better than a
leading order calculation because of its reduced uncertainty from
uncalculated terms in the perturbative expansion. Some of the
uncalculated terms contain logarithms of the renormalization scale.
Thus one way to crudely estimate this uncertainty is to vary the
renormalization scale by, say, a factor of two and see how the
calculated quantity responds. In Fig.~\ref{fig:f3vsmu}, we try this
for $f_3[\mbox{\tt NLO+PS+Had}]$ at $y_{\rm cut} = 0.05$. We vary the
renormalization scale $\mu$ in the perturbative part of the
calculation while keeping the {\tt Pythia} parameters unchanged. We
see that $f_3[\mbox{\tt NLO+PS+Had}]$ changes by about 5\% when we
increase or decrease $\mu$ by a factor 2. This is quite a small change
that, we suspect, underestimates the uncertainty, since the difference 
between $f_3[\mbox{\tt NLO+PS+Had}]$ and $f_3[\mbox{\tt NLO}]$
is 10\% . 
The $\mu$ dependence
of the NLO calculation is even smaller. By way of comparison, we vary
by a factor 2 the parameter in {\tt Pythia} that controls the scale
({\tt PARJ(81)}).  As shown in Fig.~\ref{fig:f3vsmu}, the result
changes by about 15\%, consistent with the scale variation of the pure
LO result.

\begin{figure}[t]
\vspace*{-10mm}
\includegraphics[width = 13 cm]{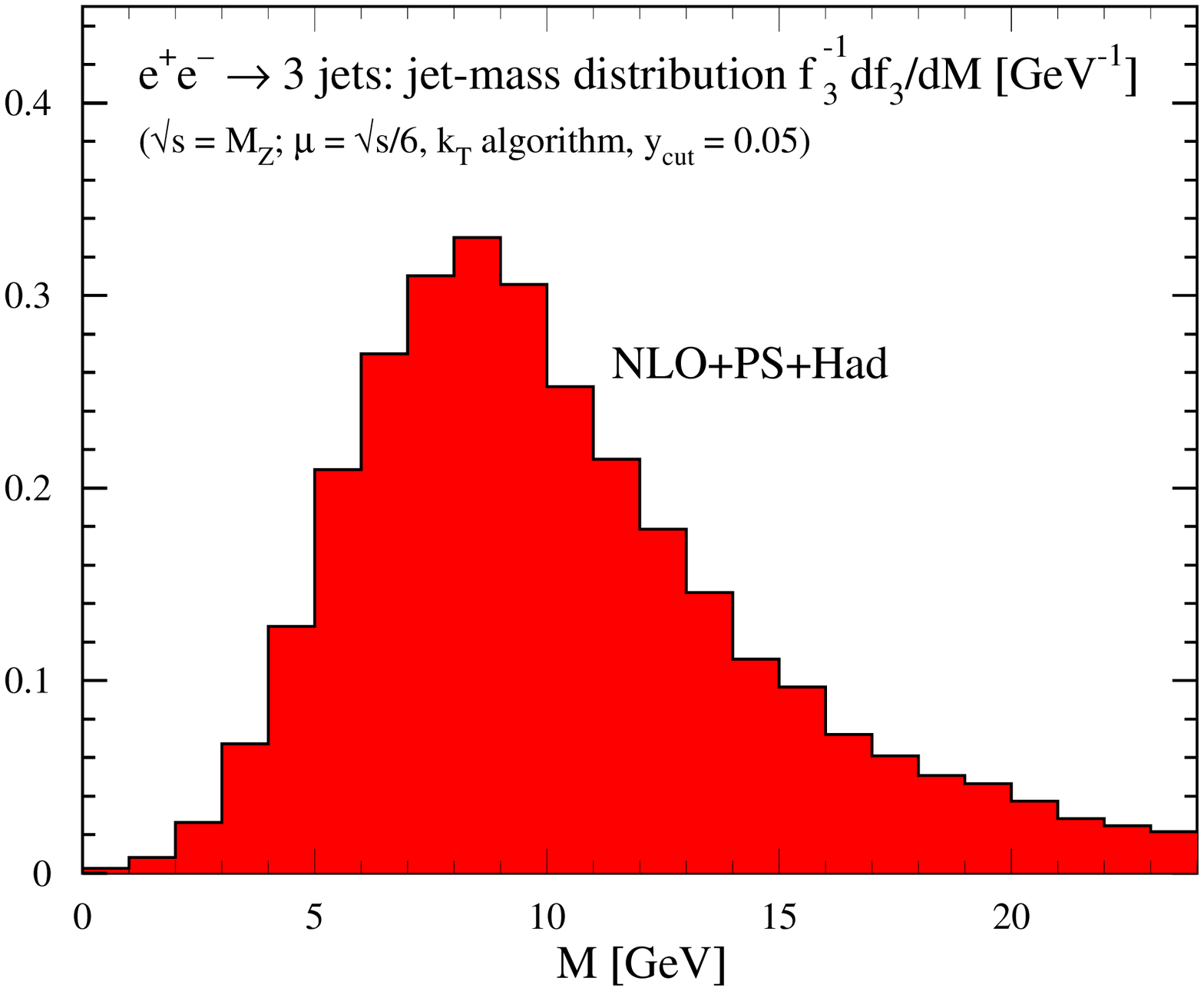}
\vspace*{-7mm}
\caption{
  Distribution of jet masses in three jet events, $f_3^{-1}\,df_3/dM$,
  in the full {\tt NLO+PS+Had} calculation using $y_{\rm cut} = 0.05$.
  The result is substantially changed from Fig.~\ref{fig:df3dmNLO} and
  is much closer to the result in Fig.~\ref{fig:df3dmPythia}. The
  calculation is defined as in Fig.~\ref{fig:df3dmNLO}. }
\label{fig:df3dmFull}

\vspace*{-7mm}
\includegraphics[width = 13 cm]{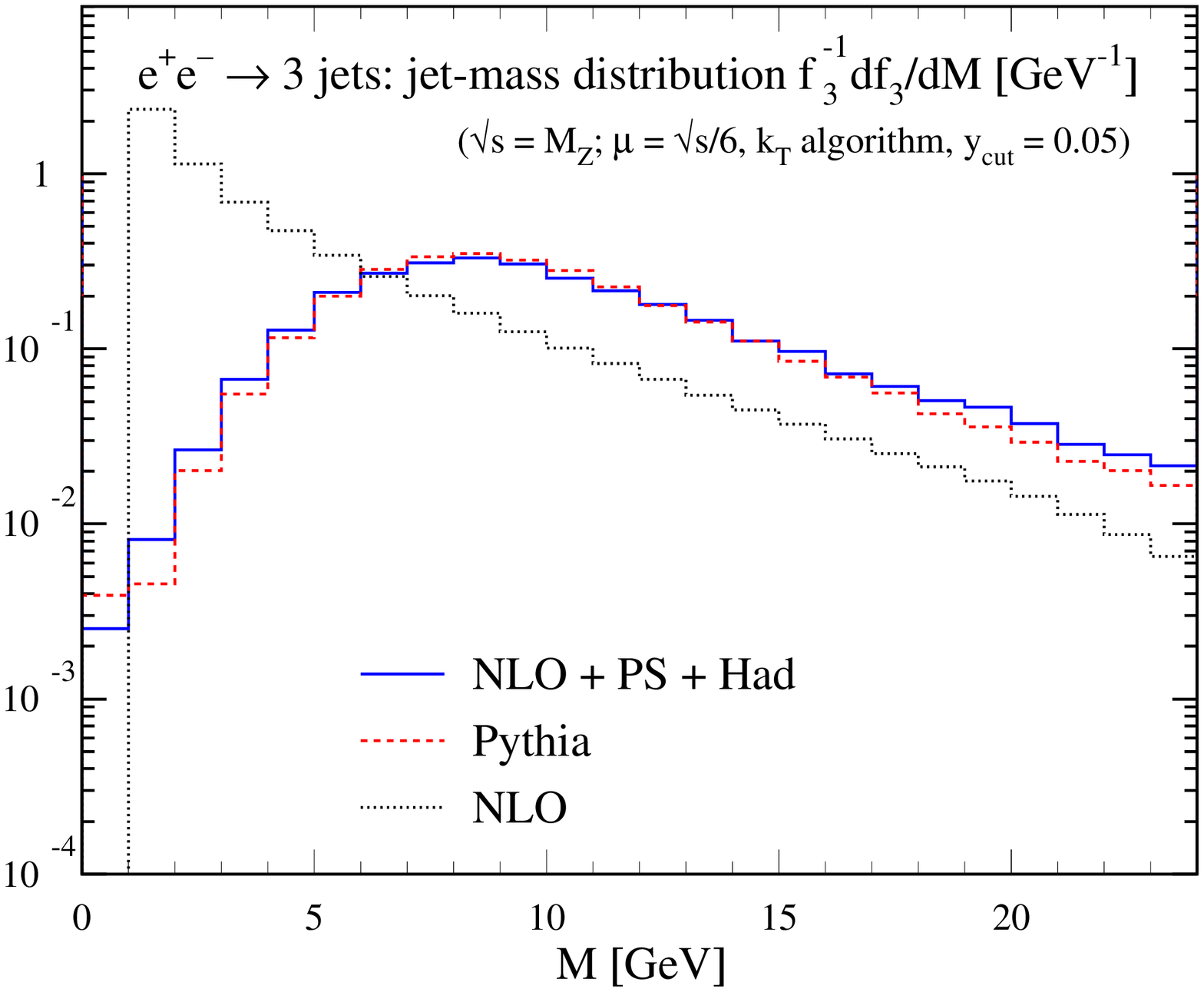}
\vspace*{-7mm}
\caption{
  Distribution of jet masses in three jet events, $f_3^{-1}\,df_3/dM$,
  in the full {\tt NLO+PS+Had} calculation, in the {\tt Pythia}
  calculation and in the pure {\tt NLO} calculation.  The calculations
  are defined as in Figs.~\ref{fig:df3dmNLO} and
  \ref{fig:df3dmPythia}. The {\tt NLO} distribution is negative in the
  first mass bin, although this is not visible in a semilog plot.}
\label{fig:df3dmcompare1}
\end{figure}

Now we turn to the distribution of jet masses in three jet events,
$f_3^{-1}\,df_3/dM$.  We recall from Fig.~\ref{fig:df3dmNLO} that
$f_3^{-1}\,df_3[{\tt NLO}]/dM$ is completely unrealistic in the small
mass region. In Fig.~\ref{fig:df3dmFull}, we plot $f_3^{-1}\,
df_3[\mbox{\tt NLO+PS+Had}]/dM$ versus $M$. Both the unbounded
increase in the cross section as $M \to 0$ and the large negative
contribution at $M = 0$ are gone. In fact, the distribution looks a
lot like the distribution produced by {\tt Pythia},
Fig.~\ref{fig:df3dmPythia}, as can be seen from the direct comparison
between $f_3^{-1}\,df_3[\mbox{\tt NLO+PS+Had}]/dM$ and
$f_3^{-1}\,df_3[{\tt Pythia}]/dM$ in Fig.~\ref{fig:df3dmcompare1}. It
is worthy of notice in Fig.~\ref{fig:df3dmcompare1} that {\tt Pythia}
gets a result for $f_3^{-1}\,df_3/dM$ that is in good agreement with
$df_3[\mbox{\tt NLO+PS+Had}]/dM$ without using next-to-leading order
corrections.

\begin{figure}[htb]
\vspace*{-10mm}
\includegraphics[width = 13 cm]{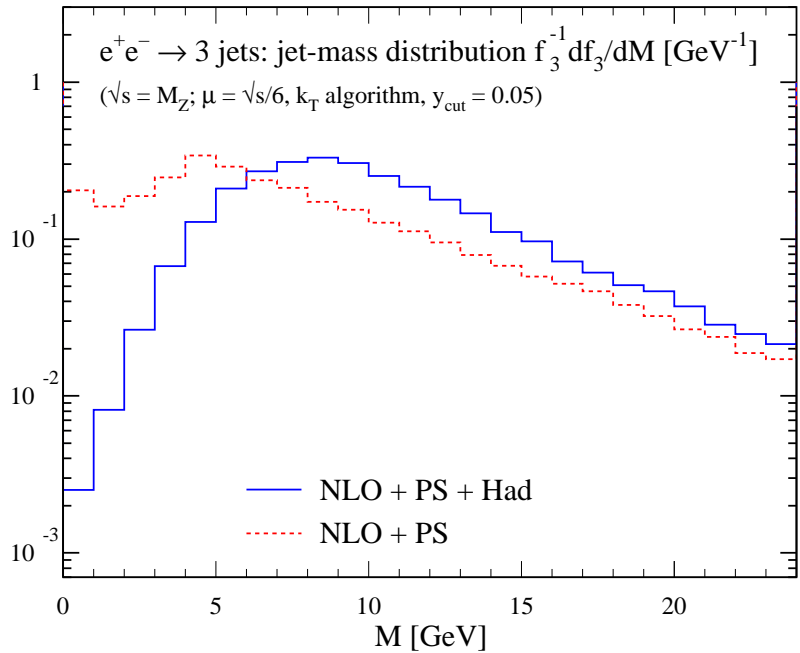}
\vspace*{-7mm}
\caption{
  Comparison of $f_3^{-1}\,df_3[\mbox{\tt NLO+PS+Had}]/dM$ to the same
  quantity not including hadronization, $f_3^{-1}\,df_3[\mbox{\tt
    NLO+PS}]/dM$. Hadronization increases the masses of jets. The
  calculation is defined as in Fig.~\ref{fig:df3dmNLO}.}
\label{fig:df3dmcompare2}
\end{figure}

Finally, in Fig.~\ref{fig:df3dmcompare2} we investigate the effect of
hadronization by comparing $f_3^{-1}\,df_3[\mbox{\tt NLO+PS+Had}]/dM$
to the same quantity with hadronization turned off, $df_3[\mbox{\tt
  NLO+PS}]/dM$. We see that without hadronization, both the unbounded
increase in the {\tt NLO} cross section as $M \to 0$ and the large
negative contribution at $M = 0$ are gone. However, there is still a
substantial probability to produce a jet with very small mass. With
hadronization, it is no longer probable to produce a jet with mass
less than 5 GeV. On the other hand, some 2 GeV is added to the mass of
typical large mass jets, thus boosting the cross section at any fixed
value of the jet mass above about 5~GeV.  This effect explains most of
the difference between $f_3^{-1}\,df_3[\mbox{\tt NLO+PS+Had}]/dM$ and
$f_3^{-1}\,df_3[\mbox{\tt NLO}]/dM$ at large jet masses.

\section{Parameter dependence}

The algorithm for combining an NLO calculation with showers described
in Refs.~\cite{nloshowersI,nloshowersII} incorporates some adjustable
parameters, most notably the two denoted as $\lambda_V$ and
$\lambda_{\rm soft}$. The calculated value of an observable like the
three-jet fraction is supposed to be correct to next-to-leading order
independent of the choices of values of these parameters. Thus $f_3$
should be relatively insensitive $\lambda_V$ and $\lambda_{\rm soft}$.
In this section we check whether this is so.

We begin with $\lambda_V$, which we can treat adequately with just two
paragraphs. Consider the splitting of one of the partons from a Born
graph, as represented by one of the square vertices in
Fig.~\ref{bornshower}. We call this a primary splitting. In a primary
splitting, there is an integration over the virtuality $\bar q^2$ of
the pair of daughter partons. In principle it is possible to use an
integration range $0 < \bar q^2 < \infty$, since there is an automatic
cutoff imposed by the kinematics of the calculation. However, one can
impose a cut $\bar q^2 < \lambda_V\,\vec q^{\,2}$, where $\vec q$ is
the three-momentum carried by the pair of daughter partons.  Taking
$\lambda_V < \infty$ changes the result from the Born graphs, but
there is a compensating adjustment in the order $\alpha_s^{B+1}$
terms.  One might imagine that taking $\lambda_V < \infty$ is
worthwhile because the parton splitting approximation with its Sudakov
suppression, which is sensible at small virtualities, is then used
only at small or moderate virtuality, while we use ordinary fixed
order perturbation theory in the large virtuality region. The default
value in the program is $\lambda_V = 1$.
We have calculated $f_3$ at $y_{\rm cut} = 0.05$ and $\sqrt s = M_Z$
for a range of values of $\lambda_V$. We find that taking $\lambda_V <
0.5$ makes $f_3[\mbox{\tt NLO+PS+Had}]$ too big. However, in the range
$0.5 < \lambda_V < \infty$, $f_3$ is independent of $\lambda_V$ to
within 10\%.  One might simply set $\lambda_V \to \infty$, thus
eliminating it from the algorithm, except that very large values of
$\lambda_V$ allow contributions from the region of finite $\bar q^2$
with $\vec q^{\,2} \to 0$. There is a singularity here (an artifact of
the use of Coulomb gauge in the Feynman diagrams), producing
undesirable fluctuations in the results.

We examine next the parameter denoted as $\lambda_{\rm soft}$ in
Refs.~\cite{nloshowersI,nloshowersII}. Consider the soft gluon emitted
from the antenna produced by the three partons emitted in a Born
graph, which is represented in Fig.~\ref{bornshower} as a gluon
emitted from the outgoing partons as a whole. The approximations used
for the gluon emission are valid only when the gluon momentum is
small. For this reason we impose a limit on the energy of the emitted
gluon,
\begin{equation}
E < M_{\rm soft} \equiv
\lambda_{\rm soft} \sqrt s_0\, (1-t_0)
\;,
\end{equation}
where $\sqrt s_0$ is the c.m.\ energy of the three parton final state
at the Born level (before showering) and $t_0$ is the thrust value
associated with this state. The default value of $\lambda_{\rm soft}$
is 1/3. Changing $\lambda_{\rm soft}$ changes the contribution from
the Born graphs, but there is a compensating change in the order
$\alpha_s^{B+1}$ graphs, so that the net change is of order
$\alpha_s^{B+2}$.

\begin{figure}[htb]
\vspace*{-12mm}
\includegraphics[width = 13 cm]{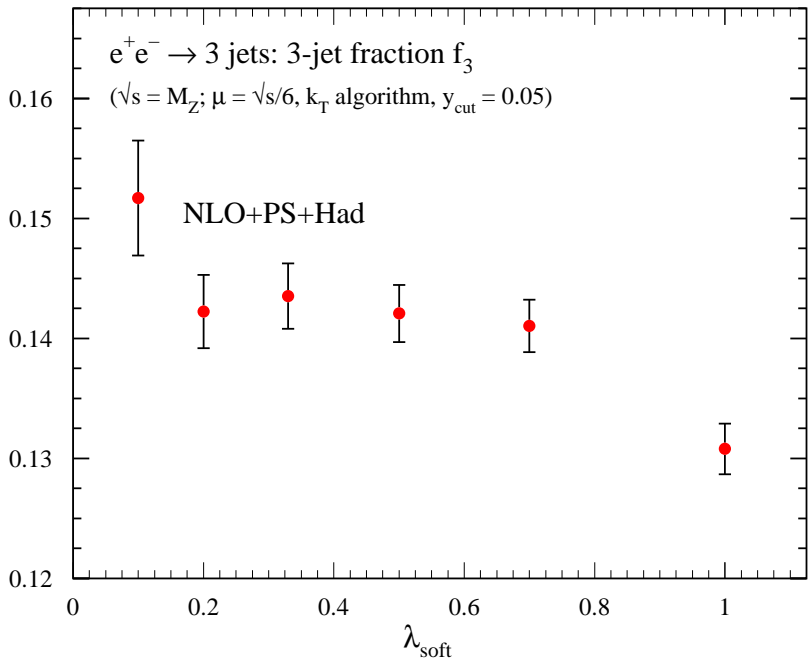}
\vspace*{-7mm}
\caption{Dependence of $f_3[\mbox{\tt NLO+PS+Had}]$ on the soft gluon
  cutoff parameter $\lambda_{\rm soft}$. The calculation is defined as
  in Fig.~\ref{fig:df3dmNLO}.}
\label{fig:lambdasoft1}
\end{figure}

In Fig.~\ref{fig:lambdasoft1} we show how the calculated three-jet
fraction depends on $\lambda_{\rm soft}$. 
We see that varying $\lambda_{\rm soft}$ over a substantial range,
$0.1 < \lambda_{\rm soft} < 1$, results in only a 10\% change in the
answer. We also see that, if we make $\lambda_{\rm soft}$ much smaller
than 1/3, the statistical uncertainty in the numerical integration
increases. The reason is that in the $\alpha_s^{B+1}$
graphs there are subtraction terms that remove the leading soft gluon
singularities and the integration range covered by these subtraction
is $|\vec k| < M_{\rm soft}$.  Effectively this inserts a cutoff
$|\vec k| > M_{\rm soft}$ in the integrations over gluon momenta in
the $\alpha_s^{B+1}$ graphs. With a very small value of $\lambda_{\rm
  soft}$, the cutoff is removed. Then real emission graphs cancel
virtual loop graphs after integration, but we encounter large positive
and negative values point by point in the integration.

\begin{figure}[htb]
\vspace*{-10mm}
\includegraphics[width = 13 cm]{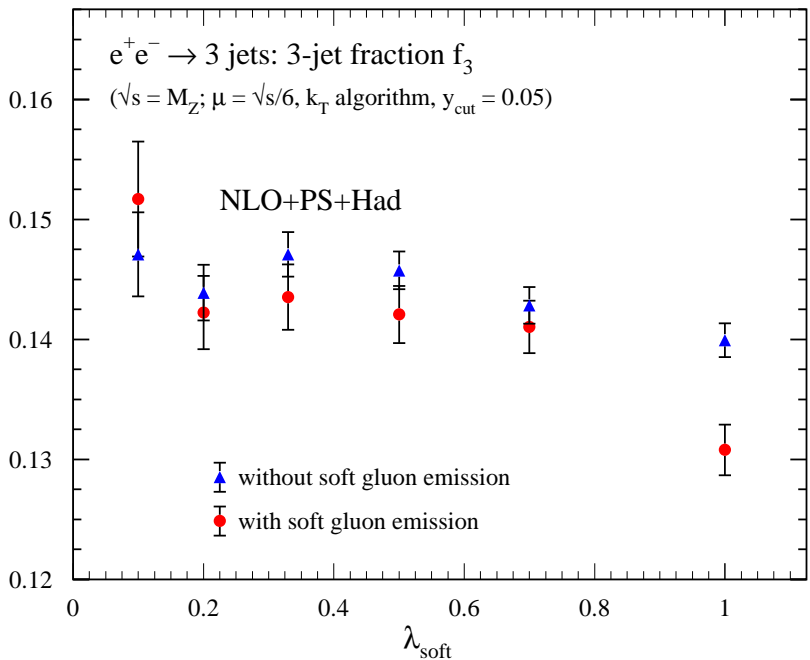}
\vspace*{-7mm}
\caption{Dependence of $f_3$ on the soft gluon
  cutoff parameter $\lambda_{\rm soft}$ if the soft gluon emission
  from Born graphs is omitted. Also shown is $f_3[\mbox{\tt
    NLO+PS+Had}]$ from Fig.~\ref{fig:lambdasoft1}. The calculation is
  defined as in Fig.~\ref{fig:df3dmNLO}.}
\label{fig:lambdasoft2}
\end{figure}

There is another approach that one might have taken. One might leave
the subtraction terms in the order $\alpha_s^{B+1}$ graphs but simply
omit the soft gluon emission from the Born graphs. This is not valid
at next-to-leading order when $\lambda_{\rm soft}$ is finite, but it
is allowed if $\lambda_{\rm soft}$ is small enough. The reason is that
the soft gluon emission from the Born graphs does not affect the
measurement function as long as the gluon momentum is very small and
the measurement function is infrared safe.  Thus leaving out the soft
gluon emission from the Born graphs is allowed if we check the
dependence of the calculated quantity on $\lambda_{\rm soft}$ in order
to see how small is ``small enough.'' This is similar to the procedure
adopted in Refs.~\cite{FrixioneWebberI, FrixioneWebberII,
  FrixioneWebberIII}, where the corresponding cutoff parameter is
called $\beta$. We have tried this for the three-jet fraction and show
the results in Fig.~\ref{fig:lambdasoft2}. What we find is that even
$\lambda_{\rm soft} = 1$ is ``small enough.'' That is, we need the
subtraction terms, or some other soft gluon cutoff, in the
$\alpha_s^{B+1}$ graphs in order for the numerical integrations to
work, but the three-jet fraction is so insensitive to soft gluons that
one could do without the soft gluon emission that compensates for the
net effect of the subtraction terms. This finding helps to justify the
approach in Refs.~\cite{FrixioneWebberI, FrixioneWebberII,
  FrixioneWebberIII}.

\section{Conclusions}

We have presented results from a method for adding parton showers to
next-to-leading order calculations in QCD when the Born process
involves massless strongly interacting partons. Specifically, the
process investigated is $e^+e^- \to 3\ {\it jets}$. We start with an
algorithm \cite{nloshowersI,nloshowersII} for coupling the NLO
calculation to the first, hardest step in showering. Here the problem
is to include the parton splittings that are part of the NLO
calculation and the parton splittings that are part of the first step
of showering without double counting. For our numerical results we
have matched the NLO calculation to the standard Monte Carlo event
generator {\tt Pythia}, although the method is more general and there
is no limitation in principle to using another event generator. We
have arranged for {\tt Pythia} to be able to accept the partly
developed shower and perform the rest of the showering plus
hadronization. With an NLO program working together with {\tt Pythia},
one can accomplish two things at once: calculate infrared-safe three
jet quantities like the three jet fraction $f_3$ with the reduced
theoretical error characteristic of a NLO calculation and get a
sensible description of the inner structure of the three jets.

Future NLO-MC hybrid programs will likely be more sophisticated than
this one and less tied to a particular Monte Carlo program than that of
Ref.~\cite{FrixioneWebberI, FrixioneWebberII, FrixioneWebberIII}. They may
employ the refined and simplified version of the matching algorithm
presented in Ref.~\cite{Nagy:2005aa}. This newer algorithm is based on a
commonly used subtraction formalism for performing NLO calculations, and
it provides more flexibility by allowing to switch from a three-jet
description to a two-jet description for events that are close to a
two-jet configuration.

\begin{acknowledgments}

We thank Z.~Nagy for advice. This work was supported in part by the
U.S.~Department of Energy.

\end{acknowledgments}


\end{document}